# A continuous integration and web framework in support of the ATLAS Publication Process

**Juan Pedro Araque Espinosa**[f] **Gabriel Baldi Levcovitz**[a] **Riccardo-Maria Bianchi**[i] **Ian Brock**[d] **Tancredi Carli**[b] **Nuno Filipe Castro**[f,g] **Alessandra Ciocio**[h] **Maurizio Colautti**[e] **Ana Carolina Da Silva Menezes**[a] **Gabriel De Oliveira da Fonseca**[a] **Leandro Domingues Macedo Alves**[e] **Andreas Hoecker**[b] **Bruno Lange Ramos**[a] **Gabriela Lemos Lúcidi Pinhão**[a,f] **Carmen Maidantchik**[a] **Fairouz Malek**[c] **Robert McPherson**[j] **Gianluca Picco**[e] **Marcelo Teixeira Dos Santos**[a]

[a] *Universidade Federal do Rio De Janeiro COPPE/EE/IF, Rio de Janeiro, Brazil*
[b] *CERN*
[c] *LPSC, Université Grenoble Alpes, CNRS/IN2P3, Grenoble, France*
[d] *Physikalisches Institut, Universität Bonn, Bonn, Germany*
[e] *Dipartimento Politecnico di Ingegneria e Architettura, Università di Udine, Udine, Italy*
[f] *Laboratório de Instrumentação e Física Experimental de Partículas, Lisbon, Portugal*
[g] *Departamento de Física, Escola de Ciências, Universidade do Minho, Braga, Portugal*
[h] *Lawrence Berkeley National Laboratory and University of California, Berkeley, USA*
[i] *Department of Physics and Astronomy, University of Pittsburgh, Pittsburgh PA, USA.*
[j] *Department of Physics and Astronomy, University of Victoria, Victoria BC, Canada*

*E-mail:* `fmalek@lpsc.in2p3.fr`

Abstract: The ATLAS collaboration defines methods, establishes procedures, and organises advisory groups to manage the publication processes of scientific papers, conference papers, and public notes. All stages are managed through web systems, computing programs, and tools that are designed and developed by the collaboration. A framework called FENCE is integrated into the CERN GitLab software repository, to automatically configure workspaces where each analysis can be documented by the analysis team and managed by the relevant coordinators. Continuous integration is used to guide the writers in applying consistent and correct formatting when preparing papers to be submitted to scientific journals. Additional software assures the correctness of other aspects of each paper, such as the lists of collaboration authors, funding agencies, and foundations. The framework and the workflow therein provide automatic and easy support to the researchers and facilitates each phase of the publication process, allowing authors to focus on the article contents. The framework and its integration with the most up to date and efficient tools has consequently provided a more professional and efficient automatized work environment to the whole collaboration.

# Contents

# 1 Introduction

The ATLAS Physics and Committees Office (also known as the Physics Office, or PO) is one of the ATLAS collaboration's [1] executive committees. It is composed of physicists and engineers performing tasks connected to the continuous support of committees and groups including the ATLAS management, the physics coordinators, the publication committee, analysis group conveners, the authorship committee, the speakers committee, and many others. The PO also provides assistance to any member of the ATLAS collaboration, by, for example, facilitating membership, authorship, paper submission to the arXiv and journals, and the review of talks and posters for national and regional meetings.

The PO supports the development of several tools including those used to manage physics analyses, prepare and submit papers, distribute detector performance documents, and track conference proceedings. It uses web-based systems to track the metadata connected with analyses, version control for editing documents, and author lists. PO members are available to guide users in understanding the tools and also assist with other daily tasks to lower the load on each member of the collaboration.

The ATLAS collaboration has an extensive organizational structure for work on detector maintenance and operation, computing, data analysis, and scientific publication and outreach. Collaborative tools are needed to provide efficient communication among collaborators and straightforward interaction with the journals, the institutions, and the funding agencies.

This paper focuses on the infrastructure for managing analysis and papers, especially its most recent developments, which were launched in fall 2017. Due to the phasing out of the use of Apache Subversion [2], a new system with document version control functionality was built using the FENCE framework. The new system is based on Git [3] and the associated CERN GitLab[1] code repository hosting platform [4]. This system is now used to handle any analysis or document type, for internal use or for publication. The FENCE framework is used not only for ATLAS document handling, but for the organization of information about other entities including members, institutes, appointments, equipment, talks, and conferences. It is also used by the ALICE and LHCb experiments for similar purposes.

Many business workflow management software packages already exist in the market, such as Monday [5] and Kissflow [6]. Those tools offer similar features like workflow definition, tracking and email notifications, but they are meant for automatizing any type of business workflow, whereas this work was implemented to satisfy solely the ATLAS publication process. Therefore, in this project, it was decided that a new and personalized solution should be developed, adapted to the LHC and physics community needs. The main reason for this decision was that integration with other internal systems and Application Program Interfaces (APIs) had already been developed using the FENCE framework, and using a different solution would have required significant effort to rebuilding these connections.

This paper is organized as follows. Section 2 describes the ATLAS publication process in general. The FENCE framework is described in Section 3, and its organization of the early stages of an analysis and integration with GitLab is explained in Section 4. The ATLAS GitLab area for editing documents and submitting papers to the journals, PO-GitLab, is described in Section 5. A description of the main tools used to support the collaboration author list and the acknowledgements of funding agencies and foundations is given in Section 6. A summary is given in Section 7.

---

[1]Gitlab is a Git-repository manager platform used by the ATLAS collaboration platform to host its Git repositories.

## 2   ATLAS publication process strategy

The ATLAS experiment supports a wide physics program to explore the fundamental nature of matter and the associated forces. To do so, it makes use of the Large Hadron Collider (LHC), which collides protons at almost the speed of light and a center-of-mass energy of 13 TeV. To carry out such a physics program, physicists need software and graphical tools to analyze the data and compare them to theoretical models. Once a data analysis is finished, the group of researchers start developing one or more documents that will be published to report their results.

ATLAS is organized into several Physics and Combined Performance working groups and subgroups, as well as several Projects (e.g. sub-detector projects) and Activities (e.g. the Trigger or Computing activity). These groups are coordinated by conveners appointed by the collaboration for typically two years.

The publication process facilitates the communication among editors and the ATLAS collaboration as a whole, and also supports the editing and review. The publications are categorized in types:

- PAPER: these are publications in peer-review journals, based on collision data analyses and/or detector projects;

- PUB notes: these are unpublished public documents classified as notes; they sometimes use only simulated data;

- PROC and CONF notes: these are conference proceedings and notes, respectively, containing preliminary results which are shown at conferences;

- INT: these are internal notes or technical documents;

- PLOT: these are plots that can be used along with the above-mentioned documents.

The most complex workflow involves the approval and submission of a PAPER. Therefore, this workflow is explained in detail below. The workflow is illustrated in Figures 1 to 3 and is composed of four phases: Phase 0, Phase 1, Phase 2 and Submission. The other document workflows normally comprise a subset of the steps present in a PAPER workflow, and thus their details are omitted.

Phase 0 represents the launch of the workflow. There, the responsible groups are constituted; the Analysis Team (AT), a group of editors of an Analysis document, is formed; the group and subgroup conveners in charge of overseeing the analysis are identified; and, finally, an Editorial Board (EB) is set up to support the analysis review. At this stage, the main physics goals of the publication are formalized, and the proposal goes through an approval process. The first integration with GitLab follows immediately, when a dedicated repository is created, based on a default document template defined for the ATLAS collaboration. This new repository contains all the necessary files to make the continuous integration (described in Section 5) work properly during the draft development. After that, the AT starts writing the first paper draft and the supporting documents.

The EB has the responsibility to enforce the implementation of ATLAS policies, review the analysis and corresponding documentation, and decide if the analysis is worth publishing in the proposed form. The Publication Committee (PubComm) chair may be consulted in decisions made during this process. The editors and conveners are also involved to support the definition of the final format of the paper – a letter or an article – and the target journal.

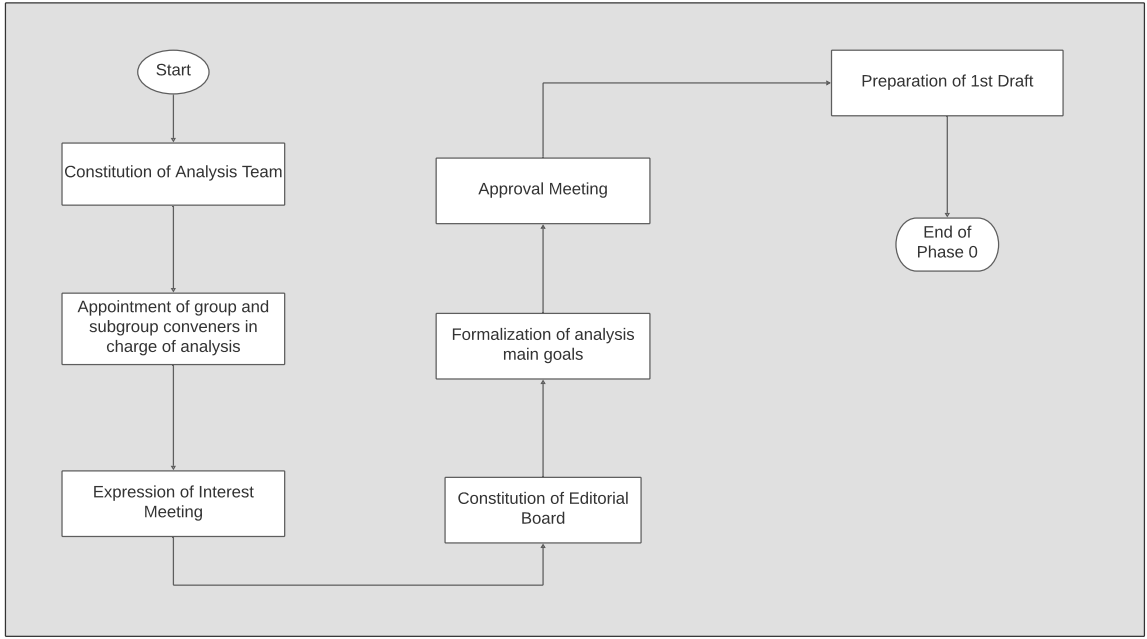

**Figure 1**: A representation of the initial Phase 0 of a PAPER workflow, when the supporting groups are formed and the analysis goals defined.

The next two phases, Phase 1 and Phase 2, consist of the validation of the draft document and its first and second circulation to the ATLAS collaboration, respectively. A paper approval meeting is held after the first circulation, in order to discuss the feedback given by the collaboration and the EB. After the second circulation, there is a paper closure meeting closing the document for further editing. Once the second circulation has passed, the document is sent to the PubComm chair for a final sign-off. The ATLAS Spokesperson (SP) is ultimately responsible for the scientific quality of the results from the ATLAS collaboration and makes a final review of each paper before the Submission phase starts.

The final draft can be signed off by the SP or their delegate. When the SP has signed off, the Phase 2 workflow is complete. A message to the PO is generated to inform it that a new document is ready to submit. This marks the start of the Submission phase.

An internal CERN preprint is produced and the final publication title is defined. Further on, a public web page is setup containing the figures and tables in the paper, as well as additional supporting figures and tables, and the PO officers proceed with the submission to the arXiv. At this stage, the proposed peer-review journal is contacted to receive the document for review. The PO officers are responsible for communication with the journal during all the steps (referee reports and proofs) throughout the whole Submission workflow. The referee reports are reviewed together with the AT and the EB, and a formal answer to each question raised is prepared to be sent back to the journal.

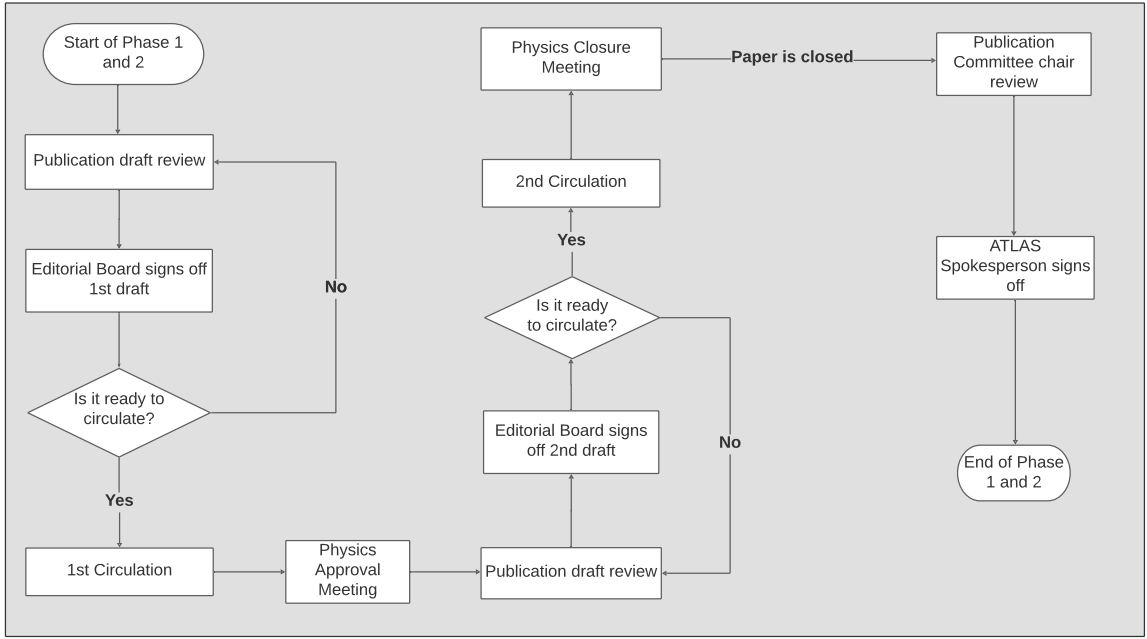

**Figure 2**: A representation of the Phases 1 and 2 of a PAPER workflow. These two phases include the collaboration review of the document being produced, as well as the final sign-off by the ATLAS Spokesperson.

Author lists and acknowledgements are both generated and handled through the FENCE framework, described in Section 3. Their production is described in detail in Section 6.1. Before the final publication and after the refereed review and acceptance by the journal, proofs are sent to the collaboration for a last check. While the editors proofread the content of the paper within a short period of time, usually two days, the PO officers check whether the authors and their affiliations have been appropriately handled by the journal, through comparison with the original files sent to them. This check is performed in large part automatically using a tool called the Proof Checker, which is described in Section 6.2. If everything is correct, the document is published online. The journal references and any previous replacements (e.g. for erratum submission and acceptance) are finally incorporated into the record. This is the last step of the workflow, which closes the procedure and makes available the references to the publication on the arXiv, public web pages, and the INSPIRE-HEP database [7].

ATLAS members interact with this workflow through web-based systems developed using the FENCE framework, which is described in Section 3. The technical implementations for each phase, with a focus on Phase 0, as well as the integration with GitLab are described in Section 4. The metadata filled in any of the Phases are exported to web sites to display the necessary information, including the Public Results pages. Some of the metadata are also used internally by the collaboration to monitor the journal submission process or related activities. The GitLab Continuous Integration (CI) tools, which are explained in Section 5, allow validation of the document drafts and preparation of the appropriate ready-to-go tarball, a compressed set of files, containing the full LaTeX [8] resources and files for the submission to the arXiv and peer-review journals.

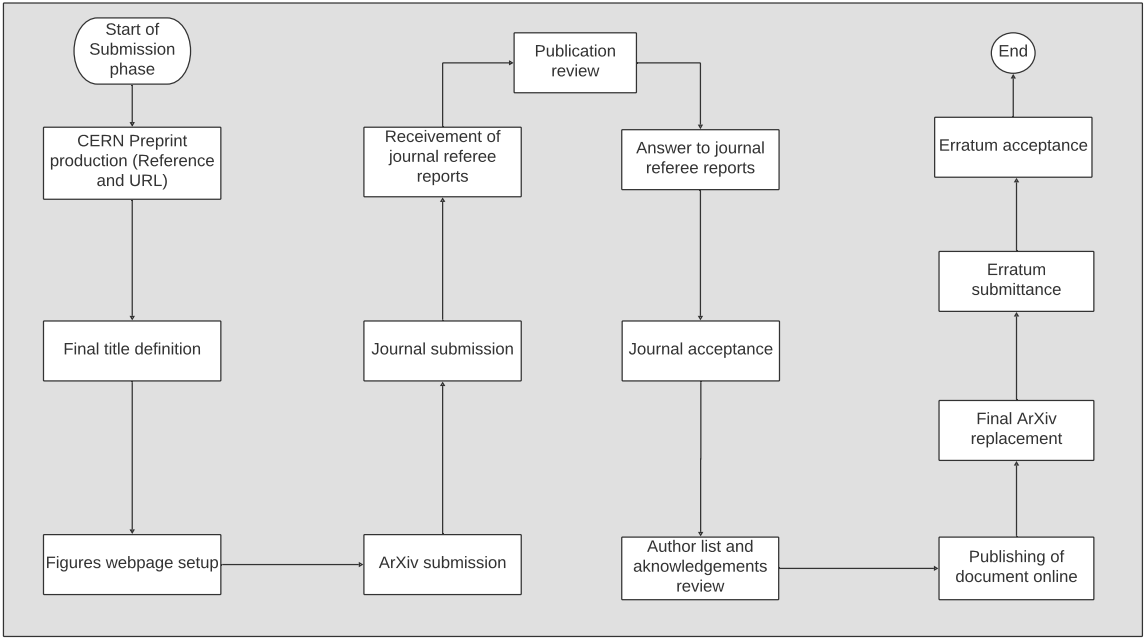

**Figure 3**: A representation of the Submission phase of a PAPER workflow.

## 3 The FENCE framework

FENCE is an object-oriented PHP [9] framework designed for the development of web applications. It encompasses the concepts of encapsulation, data abstraction, polymorphism, and inheritance. FENCE uses an Oracle [10] database (DB) to store the data fetched and displayed in its interfaces. Oracle is the default DB management system used, although with some development effort, other relational database services such as MySQL [11] and Microsoft SQL Server can be used.

### 3.1 The main concepts of FENCE

The FENCE framework is composed of a library of helper classes that are extensible program-code templates for creating objects. Any new class can be coded and added to the framework, extending its capabilities, and can then be reused in different systems. One example is the `Search` class that provides methods to create search interfaces to filter the data through predefined search attributes. Likewise, the `SuperSearch` class offers an advanced search interface where the user can build queries using logical operators. The inputs that are entered into a form can easily be added using classes such as `DateInput` and `MemberInput`, which provide a calendar-based input and a selection box with the list of all members of the collaboration, respectively.

The FENCE `Workflow` class represents a state machine. It is based on the concept of Directed Cyclic Graphs (DCG) that encompasses the relationships between objects. `Nodes` and `Edges` compose a `Workflow` that, in the case of ATLAS publication process, represent the several steps and tasks associated to it. Examples of these tasks are triggering the creation of an E-group, a mailing list solution provided by CERN, or performing an update of a GitLab repository.

One functionality for workflows is to send out e-mail notifications based on its state transition events. In FENCE, this is handled by the `Messenger` class, which provides support for sending notifications. The messages are configured using templates stored in JSON files, a lightweight format for storing and transporting data, which contain the notifications parameters, such as recipients, subject and body. Thus, notifications can be integrated into a workflow and sent out dynamically, with variable parameters according to the specific context.

Since systems built using FENCE are essentially information systems, accessing data is the preponderant task. For this, an infrastructure of Models, Builders, and Factories (MBF) was implemented. Instead of writing a plain SQL query, the MBF engine enables fetching data via orders, which are lists of the desired information. These lists are stored in JSON files and gather the properties from those available in a factory inventory that should be built.

## 3.2 Configuration files in FENCE

The FENCE framework is based on configuration files that provide the necessary parameters and properties to build interfaces. The main goal of this infrastructure is to simplify many aspects of web system requirements. The configuration is stored in JSON files. Since those can be easily converted into structured objects, it enables the definition of properties within specific contexts. One usage of this feature is to set up access permissions to a given interface, by defining the group membership required for a user to be allowed to proceed, as explained in Section 3.3.

One of the benefits of using configuration files is that classes with numerous arguments can be instantiated in a cleaner way, with just a configuration file path as a constructor argument. This practice gives the opportunity to add more context behind configuration values that would, otherwise, be simple scalars. Concrete cases employed in most of the FENCE system interfaces are field definitions, including edition permission and validation callbacks, MBF orders (see Section 3.1), and workflow step definitions. In these cases, centralizing the information in a JSON file improves the code maintainability by using human-readable text to describe the context of high-level features, for example, in a web form. Additionally, it makes it easier for the code base to scale in case of disruptive changes such as moving to a different programming language, or refactoring current features. Moreover, all specifications of the interface are defined in a single JSON file, facilitating the verification of compatibility between old and new implementations.

## 3.3 Security in FENCE

Alongside the features FENCE offers, securing the information stored in the applications' databases is also a responsibility of the framework. For that, authentication and authorization layers are built-in to ensure, respectively, the authenticity of users' credentials and the availability of proper resources based on the users' permissions.

CERN provides an official Single-Sign On (SSO) authentication solution, which is used by all applications available within the CERN network. The authentication service provides a list of E-groups that a member owns or is part of in the collaboration. FENCE implements a `BaseUser` class which makes use of the information given by the SSO to control access to the application resources. Additionally, an internal validation is done to verify the registration of the authenticated user in the application database. This extra check is needed to map the authenticated user identifier to the one used internally by the FENCE applications to define relationships with other entities.

Once the authentication is done, the set of functionalities available for the user is restricted or expanded according to the roles to which they are associated. These roles can be defined either by the users' E-groups, obtained from the authentication service, or user groups defined in the application database. A `User` class encapsulates the access to the users' roles and is ultimately used to check them when building interfaces.

## 4 The analysis web-based systems with a focus on its initial phase

The need for tracking of the initial phase of an analysis, Phase 0, arose 2017, as the CERN IT department phased-out the SVN version control system and encouraged the use of Git [3] because of its decentralised characteristics, which are better suited to the reality of the ATLAS collaboration. The experiment started to use the repository platform GitLab [4] because of its continuous integration functionality, the possibility of storing repositories in private servers, and the provisioning of an API with many services.

To formalize the creation of Git repositories at the beginning of the publication writing process, the concept of Phase 0 emerged. This concept was opportunistically used to track the flow of tasks during the preliminary stage of the editorial process.

### 4.1 The main functionalities of the Phase 0 system

The Phase 0 system incorporates three main interfaces. The first is presented in Figure 4 and allows the submission of a new analysis. At this stage, ATLAS users must provide the initial description of the article or public note in order to start writing. The interface presents a web form that contains several fields, some of which are mandatory. If all fields are filled correctly, the form information is stored in the database, gathering the information that defines an analysis such as its title and reference code.

The second interface presents advanced search functionality and allows a user to define complex logical expressions as search criteria. Users can also configure the search results by grouping them by attributes, selecting the visible attributes, or saving those configurations locally for use in a future search. Search results can also be exported in comma-separated value (CSV) file format.

Finally, the publication details interface, the main interface of the system, shown in Figure 5, presents metadata and allows their editing. The interface also controls the workflow of Phase 0 activities, providing an overview of all its stages and highlighting the previous, current, and upcoming ones. A transition between Phase 0 steps triggers actions. The most common action is storing data in the database. If allowed, a user has the option of saving the data and staying at the same step by pressing the "Save" button, or saving the data and moving to the next step by pressing the "Proceed" button. When one moves forward in the workflow, the system triggers automatic alert messages that provide instructions to the person(s) responsible for the next step.

An example of a Phase 0 step is the request for the formation and first meeting of an Editorial Board (EB), which is illustrated in Figure 6. The group convener is responsible for adding the EB request meeting title, date, comments, and links. The Publication Committee (PubComm) chair is responsible for appointing the EB members and entering the date on which they are appointed. Once this information is saved in the system, the PubComm chair can proceed to the next Analysis workflow step. Subsequently, the EB E-group is automatically created, including all its members, and an email is sent to those members, informing them that they were appointed and should continue the workflow.

**Figure 4**: The analysis submission functionality in the Phase 0 system. On the left is a summary of all steps needed to complete the data submission. On the right are the fields that belong to the first step of the data submission.

## 4.2 FENCE-GitLab integration

The first interaction between FENCE and GitLab happens when a Phase 0 entry is created. A GitLab group and a first internal note repository are automatically created using a boilerplate set of standard files and configurations. The first commit to this repository is obtained from a source repository, a package containing file templates. FENCE is responsible for substituting all the necessary variables into all the file templates (e.g. automatically naming some files using the analysis reference code).

Another FENCE and GitLab integration process is executed when Phase 0 is finished or is skipped, thus proceeding to Phase 1. FENCE automatically creates a publication repository, setting all the configuration elements that are needed. That means that the creation and the configuration of the repositories holding the documentation is done without any input or intervention from the editors, allowing for a streamlined process.

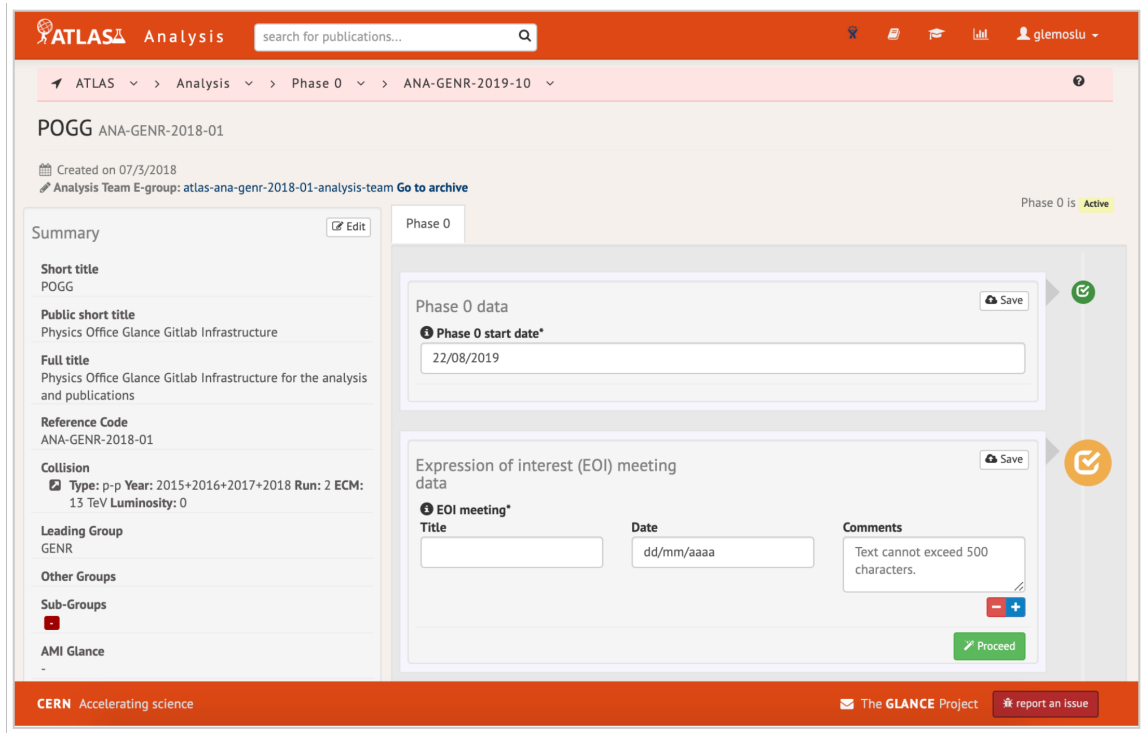

**Figure 5**: The main interface of the Phase 0 system. On the left is a summary with the most important information about a publication. On the right are the steps corresponding to the Phase 0 workflow.

FENCE and GitLab also interact while handling the author list of a publication. At first circulation, the author list is created according to its reference date in various formats (including xml and `tex`). It then pushes the files through the GitLab API so they are stored in the existing publication repository. Upon first circulation, the files are added to GitLab, while on subsequent circulations, they are simply updated.

## 5   PO-GitLab and CI tools

The ATLAS Physics Office GitLab tools (PO-GitLab) simplify the publication process of ATLAS documents by using the features provided by the CERN GitLab platform.

Previously, the publication workflow involved a heavy email exchange between ATLAS editors and the Physics Office in order to ensure that ATLAS rules were being followed up to submission of the paper to the arXiv and any peer-review journal. This approach led, usually, to modifications by different parties (officers and editors), which were sometimes incorrect or conflicting and which slowed the publication process down. Due to the uniform and repetitive nature of the tasks required to submit a publication, an automatic tool was favored.

Three main tasks are handled by the GitLab integration tools throughout the submission phase: the automatic creation of GitLab repositories; the continuous verification of technical rules by the GitLab Continuous Integration (CI) tools; and the automatic processing of the document itself for submission. These tasks are presented at detail in this section.

**Figure 6**: Screenshot of the Editorial Board request meeting and formation step in the FENCE Phase 0 system.

### 5.1 GitLab structure to organize analysis groups and repositories

A hierarchy of groups were created in GitLab to reflect the current ATLAS group organization. Each Physics or Combined Performance group, System Detector Project, and Activity is labelled as a category with four letters in the FENCE systems. For example, the Top Quark physics group is TOPQ, and the Electron/Photon Combined Performance group is EGAM. At the beginning of Phase 0, Git repositories are created in GitLab in the area corresponding to the lead group of the analysis. The identifier of a Phase 0 entry is labelled ANA-GROUP-YEAR-NN, where GROUP can be, for example TOPQ or EGAM, while YEAR is the year the document was created and NN is a two-digit counter. For instance, ANA-SUSY-2019-04 represents the fourth analysis created in the SUSY group in 2019.

In GitLab, the following labels are adopted:

- ANA-GROUP-YEAR-NN-INTn for internal note repository,

- ANA-GROUP-YEAR-NN-PAPER for a paper repository.

The other publication categories (e.g. PUB and CONF) are also included. For instance, in the Higgs (HIGG) physics group, for a given Phase 0 analysis entry ANA-HIGG-2017-08, PO-GitLab will host ANA-HIGG-2017-08-INT1,2,...,n and ANA-HIGG-2017-08-PAPER. This is illustrated in Figure 7 where the GitLab interface is shown. ANA-HIGG-2017-08, an analysis within the HIGG group, contains for example one paper and one internal note repository, respectively ANA-HIGG-2017-08-PAPER and ANA-HIGG-2017-08-INT1.

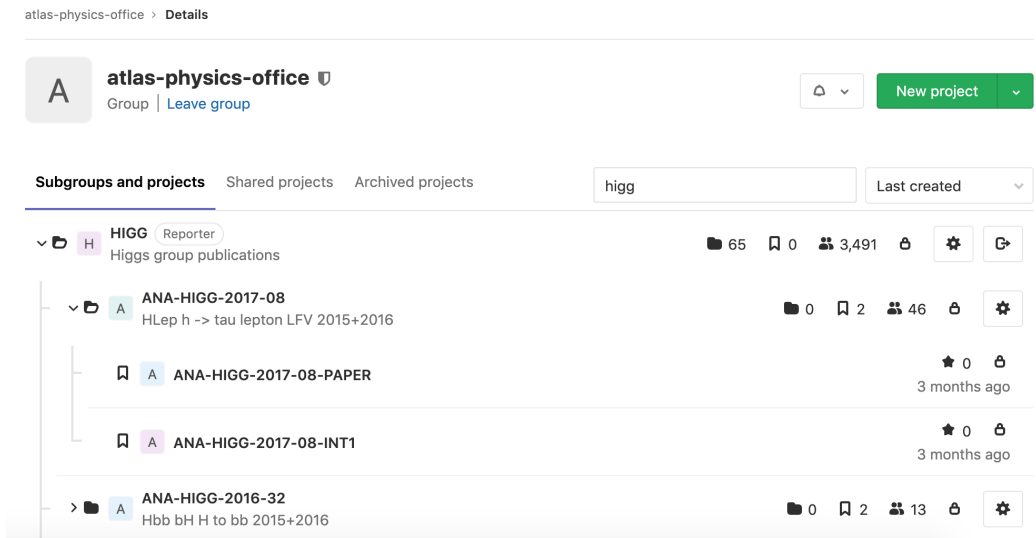

**Figure 7**: The substructure of a HIGG GitLab repository subgroup. The main admin group, atlas-physics-office, is shown at the top. The HIGG subgroup is selected, and its ANA-HIGG-2017-08 subgroup is expanded. A repository for the paper and one for the internal note (INT1) are created under ANA-HIGG-2017-08.

The automatic creation of repositories is done via communication between the FENCE framework and the GitLab API, which is explained in more detail in Section 4.2. The structure of groups also permits the definition of an LDAP (Lightweight Directory Access Protocol) based fine-grained access control provided by GitLab. It guarantees that repositories created under a given group are restricted to the appropriate set of users (group conveners, analysis participants, and PO officers).

### 5.2 Automatic document creation

The initial files and directories present in a repository created through the FENCE web interfaces are based on a template. They have their variables substituted according to requirements of the related publication in the moment of creation. This way, all created repositories contain the default documents, correctly formatted, to start writing a paper, CONF note, PUB note, or INT note. The repository is also configured with a protected branch named PO-ready, to which only members with the Maintainer role are allowed to push and merge. This special branch is used to run the final submission pipeline when the document is ready and has been reviewed by the relevant parties. The master branch is used as the main work branch, unprotected at the time of the repository creation, allowing all editors to push new commits and interact with the repository.

### 5.3 Continuous checks with GitLab CI

GitLab CI tools are designed to automatically execute a set of tasks every time a new modification is introduced into the document, which happens on every new commit pushed to the document repository. A command line tool called PO-GitLab was developed to perform a variety of checks on a given document. Those checks verify distinct aspects of an ATLAS publication, from style guidelines to internal conventions, and is implemented in Python. The application architecture is based on the isolation of each component that performs the verifications in abstractions that represent a job from the pipeline. Each job is represented as a single code path which can be executed without affecting the others, which allows for new and more complex tasks to be added continuously, and facilitates the reuse of each component for creating new pipelines. Finally, the tool is packaged up in a Docker [12] image which makes possible to run it in any environment that has Docker installed in it, beyond only the GitLab CI machines installed at CERN.

GitLab's CI offers a way to configure and organize task pipelines. A pipeline is simply a set of jobs grouped into different stages. All the jobs in the same stage are executed in parallel, while each stage is only executed after the previous one has completed. It is possible to start the jobs of one stage only if the previous ones have finished successfully, or alternatively only if one or more previous stages have failed. The tool starts a pipeline automatically based on the configuration put in place for a given repository – normally this is done every time a new commit is pushed to the repository.

Different sets of checks are performed in each step of the publication process. For editors, all work done before the paper submission (detailed in Section 5.4) is monitored by the editing-pipelines shown in Figure 8. These pipelines are triggered by any push made from Git branches whose name does not start with PO-. The special branches using the PO- prefix are tracked by the submission-pipelines when a paper is considered ready for submission to the arXiv and the peer-review journal.

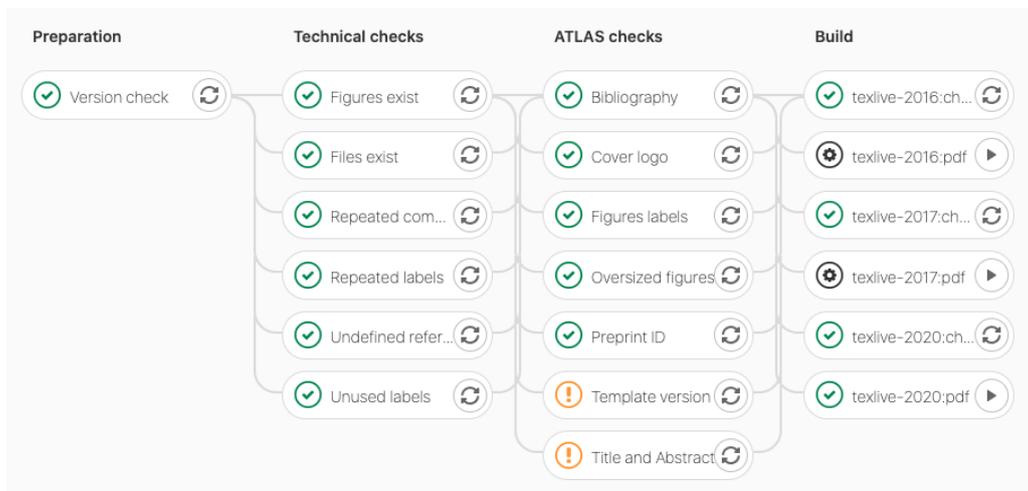

**Figure 8**: The editing-pipelines. These four stages perform checks before a publication is ready for submission, with the first stage checking the version of the PO-GitLab package itself, the second stage running checks related to LaTeX formatting, the third ensuring that the ATLAS rules are followed, and the last stage testing if the document builds correctly (without LaTeX-related errors).

## 5.4 Paper submission

The CI also produces the files required for paper submission, using dedicated pipelines similar to the editing ones. These are called submission-pipelines. They are responsible for producing the final tarball, the set of files in an archived format (`tar.gz`) that will be sent to the journal or to a public repository for building public web pages.

A protected Git branch, named PO-ready, is created by default at the time of the setup of the paper repository. When a paper is ready for submission, an editor creates a Merge Request to the PO-ready branch. When this request is accepted by a Physics Office officer, the paper submission pipelines are triggered. In addition, any branch or tag created following the pattern PO-* triggers the paper submission pipelines. These pipelines have, in addition to the previously described checks, a flattening of the LaTeX documents, grouping together all the necessary files for submission, as shown in Figure 9. The flattening goes through the following steps:

1. all the source files are merged into a single LaTeX source file;

2. all the comments in the LaTeX source file are removed;

3. all the figures are renamed following the convention required by the journals;

4. any directory structure is removed.

Tarballs suitable for submission to the arXiv and journals are created using TeX Live 2020 and 2017, respectively. The two different versions differ in their handling of the bibliography and references, and must be checked in order to avoid incompatibilities. The arXiv, at the time of this publication, requires TeX Live 2020, while some APS journals, for example, require TeX Live 2017. Other tarballs contain files with plots and tables for posting on a public web page. These tarballs are created as GitLab artifacts and can be downloaded by the corresponding editors and members of the Physics Office. In the submission tarballs, any auxiliary material (e.g. figures and tables not for submission) is not included.

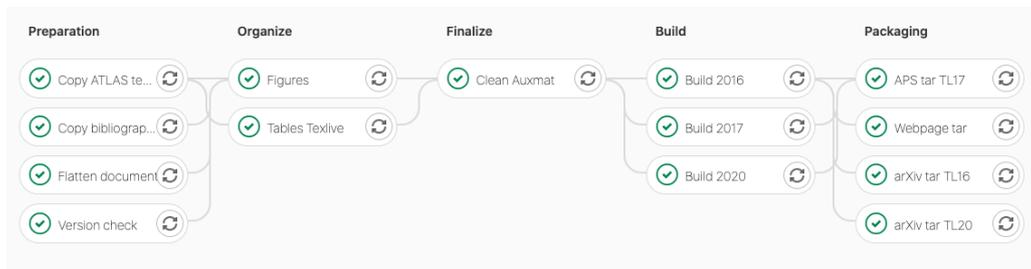

**Figure 9**: A submission-pipeline. From left to right, preparation stage jobs check the version of the CI tools, copy `bib` and `sty` files along with the flattened LaTeX document to a special folder. Then all figures and tables are renamed and labelled according to the journal's specifications. At the final stage, the flattened document is updated with the newly named figures and tables. In the last two steps, the document is built, producing the `bbl` file needed for the journal and the tarballs for the public web pages.

## 6 Author lists, acknowledgements, and the proof checker

### 6.1 Author lists and acknowledgements files

The author list, often written authorlist for convenience, is the inventory of qualified authors at a given date, which is called the reference date. Every paper has an associated list of qualified authors with a reference date that corresponds to the creation date of that list during Phase 1, just before the first circulation of the draft document to the collaboration. Between Phase 1 and Phase 2, this author list may be updated or amended. This information is stored in the ATLAS database and managed by FENCE. Figure 10 shows a portion of the full list of members, their affiliations, and the related metadata which are needed to generate the full report.

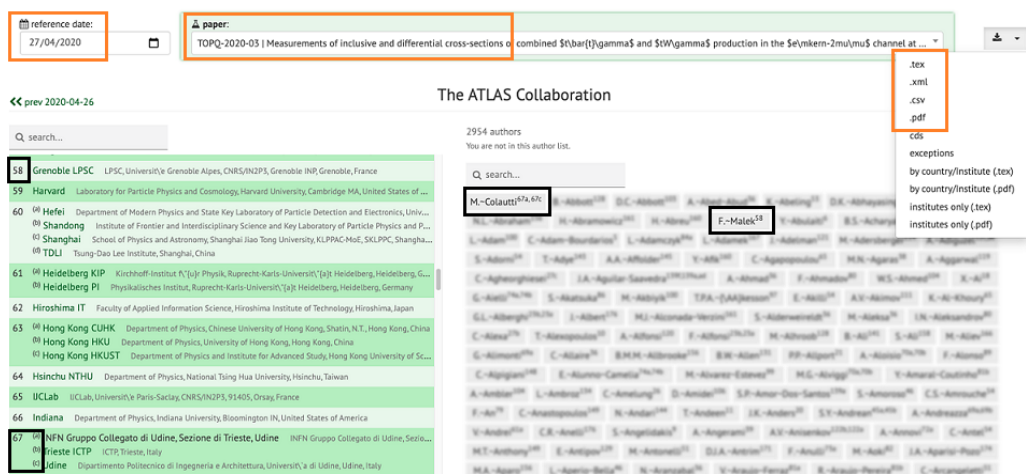

**Figure 10**: The FENCE author list generation interface. From left to right, the list of institutions used as affiliations; in the center of the screen, the authors (blurred) are listed; at the top of the page, the interface allows the users to view (orange boxes) the creation date (left) for a given paper, the title of the paper (center), and the formats available for download (`tex`, `xml`, `pdf`, etc.; top right). As an example, two authors of this paper are highlighted with their affiliations (black boxes).

The acknowledgements are incorporated in a short section that the collaboration includes in each paper to thank funding agencies for their financial support. They do not change often, but they may add or remove a funding agency or a foundation at a given date. Therefore, similarly to the author list, the acknowledgement file is built for each paper at a reference date.

Both files, the author list and the acknowledgements, are built using the FENCE framework and are automatically pushed to the appropriate GitLab repository, using the FENCE–GitLab integration (Section 4.2). Their integration into the paper is straightforward at the time of submission to a journal. FENCE provides the MBF infrastructure to retrieve the required information from the database (see Section 3.1) and build all the files.

The author list is built by the FENCE framework into an xml file. This is composed of three main blocks:

- Header: stores the paper's main information.

- Institutes: the list of institutes and their InSPIRE-HEP references.

- Authors: the list of authors and their identifying information, including names, initials, affiliations, and ORCID[2].

Most journals accept the `xml` version of the author list as input, and retrieve all the needed data from it. The complete set of author lists created for every ATLAS paper that has been submitted or published since 2009 is accessible through a web interface, called the FENCE author list interface. They are easily filtered using a search box.

The acknowledgement `tex` file is built using a standard template and is filled using the FENCE framework to retrieve the required information about the ATLAS funding agencies.

### 6.2 Proof checker functionalities

Once the author list included in the tarball has been sent to the journal, a check is made to determine whether the publisher has correctly used the provided information. This check involves a comparison of the proof `pdf` file that was sent back to the ATLAS Collaboration for review to the original `xml/tex` file. This process used to be done by hand, requiring the PO officer to check that each of the approximately 3000 authors and 200 institutes were correctly reported and matched. The proof checker is the tool provided for ATLAS to compare these two files automatically. A report of this comparison, one for every version of the proof, is available to PO officers who check the results.

The proof checker follows this process:

- retrieve the information from the `xml` file, containing the authors and their affiliations;

- extract the text from the journal's `pdf` file;

- parse the text from the `pdf` file, creating the target reference;

- compare the official reference obtained from the `xml` file with the target reference;

- create a report with the differences found between the original and the target reference;

- link the report to the main ATLAS report page.

The main difficulty with this process involves extracting the content from the `pdf` file; the text is not easily retrieved, for a variety of reasons. One is that many elements have to be identified and ignored, such as row numbers, watermarks, footers, and headings. Another reason is that words extracted from a `pdf` file don't follow a specific coding convention; the file can contain non-ASCII characters that can be extracted in many different ways. The `pdf` file can specify a predefined encoding standard to use, or provide a lookup table of differences between predefined and alternative encoding standards; for texts with uncommon Latin characters, which are routine in this kind of publication, special encoding is used, and they are translated into the Unicode standard convention for glyph representation. It is necessary to provide a ToUnicode Table, to map the codes used in the `pdf` file to Unicode symbols, where semantic information about the characters is preserved. The proof checker also has to parse all the publication text and recognize where the author list starts, where it ends, where the institute list starts and where it ends. All this is made more difficult by the fact that different publishers have different layouts and create different versions of `pdf` files. This makes the above problems not generic, but often specific to a particular publisher.

---

[2]ORCID stands for Open Researcher and Contributor ID.

After the target reference is created, the comparison looks for:

- authors that seem to be missing from the `pdf` file. Here, false positives are often due to character encoding and spaces;

- authors with inconsistent punctuation. This section points out differences between original and target references authors' first name punctuation, which can follow the rules X. or X.Y. or X.-Y. or X-Y. with or without spaces;

- institutes that seem to be missing from the `pdf` file. Here false positives are often due to the difficulties extracting and correctly encoding/decoding special glyphs in the `pdf` file format, which make the comparison fail;

- institutes with close matches. All the entries that look like the original but have some inconsistencies land in this group. Some publishers replace, for example, "USA" with "United States of America" (or vice versa). Sometimes there is a new character that doesn't corrupt the institute entry, but makes the match imperfect, for example, "Università" and "Universit'a"; this is one of the problems that often occurs due to the difficulties of extracting data automatically from a pdf file.

- mismatched authors. All the authors collaborate through one or more institutes. The link between the author and the institute is checked for consistency. This sometimes results in a false positive, because it is not always easy to extract from the `pdf` file the index number of an institute, mainly because the text coming from the journal `pdf` file also includes other elements such as line numbers of the document. For this reason an author originally assigned to institute number X can end up matched with target institute YX, because in the text extracted from the pdf the number X might be preceded by a Y line number; institute YX may not exist;

- deceased authors. In some cases, ATLAS has tagged authors as deceased but the publication fails to mark them as such, or vice versa;

- missing funding agencies, or those wrongly added by the publisher.

In early 2019, due to changes in CERN systems, the component written in Prolog which ran the comparison went out of service. This implied an urgent need for a new tool for this task. Python was chosen as the programming language for the new comparison engine.

A way to obtain the best match among all the items of an array of institutes and authors was sought, because one cannot rely on finding an author or institute in the same position of the sequences in the `xml` and `pdf` files. For this purpose the concept of Levenshtein distance [13] was applied, so that a weighted index of similarity can be obtained to decide what is matched with what, and to then effectively check for anomalies. The Levenshtein distance between two words is the minimum number of single-character edits, such as insertions, substitutions or deletions, required to change one word into the other. For example the Levenshtein distance between ATLAS and Atlassian is 4 (4 insertions); between Maurizio and Fabrizio is 2 (2 substitutions); raise and race is 2 (1 deletion, 1 substitution).

Mathematically, the Levenshtein distance $\text{lev}_{a,b}(|a|,|b|)$ between two strings a and b of length $|a|$ and $|b|$ respectively is given by:

$$
\text{lev}_{a,b}(i,j) = \begin{cases} \max(i,j) & \text{if } \min(i,j) = 0, \\ \min \begin{cases} \text{lev}_{a,b}(i-1,j) + 1 \\ \text{lev}_{a,b}(i,j-1) + 1 \\ \text{lev}_{a,b}(i-1,j-1) + 1_{(a_i \neq b_j)} \end{cases} & \text{otherwise,} \end{cases}
$$

where $1_{(a_i \neq b_j)}$ is equal to 0 when $a_i = b_j$ and equal to 1 otherwise, and $\text{lev}_{a,b}(i,j)$ is the distance between the first i characters of a and the first j characters of b.

A feature was developed to help the script evaluate as correct matches some that would not otherwise appear to be such. A list of synonyms (see Section 6.2.1) is created for every entry, author or institute, to teach the proof checker to validate similar strings when the differences are due only to problems from decoding the text from the `pdf` file. So, for instance, if author X. Nonamečič is not found in the target reference, but from the `pdf` entries an author with name X. Nonamež čiž c appears, then, as it has been previously verified that in the `pdf` file the name appears as expected, the proof checker considers it to be the same, and skips the problem. A very long list of false positives can be found in the report page as "skipped items". The list of synonyms is updated manually, but a tool, the Synonym web page, has been created to allow users to update this list themselves.

### 6.2.1 Proof checker synonyms

As previously discussed, the comparison between the `pdf` file and the `xml` file can generate false positives. To minimize the list of false positives in the report page, the proof checker includes a synonyms list that allows the comparison script to understand if the difference is a real error or another correct way to display the same information. An example of a working synonym is Physics Department, SUNY Albany, Albany **NY, United States of America** for the ATLAS original version and Physics Department, SUNY Albany, Albany, **New York, USA** for the Journal version. These differences are acceptable, since the main information is correctly displayed. All the synonym records are managed using a JSON file and are separated into institutes and authors. Institute listings are, for example:

```
{
    "id": "2",
    "original": "Department of Physics , University of Alberta , Edmonton AB , Canada"
    "synonyms": ["Department of Physics , University of Alberta , Edmonton , Alberta ,
    Canada"],
}
```

Author listings are, for example:

```
{
    "original": "A. B\\\"ub",
    "inspire": "INSPIRE-00000000",
    "foafName": "A Bub"
    "synonyms": ["A. B\u00f2b", "A. B\u00a8 b"],
}
```

Search for institute: [Udine] 🔍 (at least 3 characters)

**INSTITUTES**

-Dipartimento di Chimica, Fisica e Ambiente, Universit\`a di Udine, Udine, Italy ✏

SYNONYMS:
    Dipartimento di Chimica, Fisica e Ambiente, Universit` a di Udine, Udine, Italy
    Dipartimento Politecnico di Ingegneria e Architettura, Universitò a di Udine, Udine, Italy
    Dipartimento Politecnico di Ingegneria e Architettura , Universitò a di Udine, Udine, Italy
    Dipartimento di Chimica , Fisica e Ambiente, Universitò a di Udine, Udine, Italy
    Dipartimento di Chimica, Fisica e Ambiente, Universitò a di Udine, Udine, Italy
    Dipartimento Politecnico di Ingegneria e Architettura, Universit` a di Udine, Udine, Italy

Search for author: [Pin] 🔍 (at least 3 characters)

**AUTHORS**

-G.L.L. Pinhão ✏

SYNONYMS:
    G.L.L. Pinhao

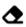

**Figure 11**: Proof checker synonyms. One can search for an item and see all the defined synonyms for institutes and authors' names.

To manage the list of proof checker synonyms, ATLAS provides a web interface that allows users to search for an existing entry and manage the recorded synonyms. Searching for an institute or author will display the list of records that match the search criteria (see Figure 11). This allows PO officers to edit the synonyms for the record. Clicking the edit icon shows a new page section where a new synonym can be added. After confirmation, this is added to the list of synonyms and is taken into account by the next run of the proof checker.

### 6.2.2 Report page

The proof checker provides a report for each paper and draft version (the journal may produce several proofs). This report is provided and stored in a JSON file and must be parsed to show the report results in a human-readable way. The report contains all the paper information and the comparison results sorted by topic, for example:

```
{
    "ref_code": "EXOT-2017-24",
    "ref_date": "2018-07-31",
    "creation_date": "29-Oct-2018",
    "publisher": "'APS'",
    "document": "doc1053",
    "filename": "LY15578_proof_v2",
    "authors_missing_skip": [...],
    "authors_missing_list": [...],
    "authors_puntuation_list": [...]
    "institutes_missing_pdf_list": [...],
    "institutes_missing_pdf_skip": [...],
    "authors_mismatched_list": [...],
    "authors_not_deceased_list": [...],
    "authors_deceased_list": [...],
    "institutes_close_matches_list": [...],
    "founding_agencies_missing": [...],
```

```
    "founding_agencies_wrong": [...]
}
```

The JSON file contains more information than is displayed; the information is reduced to allow the web page to optimize the display of the huge amount of information and to retain data for future improvements. The web page contains some hidden sections that are produced by the proof checker via the known synonyms. These can be displayed by clicking on "Skipped +". Here the page will show all the false positive results that the proof checker found in its comparison, but that are ignored after association with the synonyms.

The proof checker helps the Physics Office staff in a tedious task, but it is far from being a perfect tool. It requires continuous maintenance and updates for new cases, changes in publication layouts, and new conventions in the author lists and their format. Further improvements are planned, with the goal of minimizing the number of cases to be checked manually.

## 7 Conclusion

A suite of tools have been developed to support the publication of documents by the ATLAS collaboration. While the emphasis is on papers published in refereed journals, the new technology also supports internal documents and other public documents such as conference and public notes.

The FENCE framework is used as the backbone of the whole setup and is also used to interface the web-based tracking of the status of an analysis with the documentation in GitLab. Extensive use is made of the Continuous Integration tools available in GitLab to ensure that documents can easily be submitted to the arXiv and journals as soon as they have been approved by the collaboration.

The software solutions described in this document are now used to accompany the entirety of a physics analysis, from the expressions of interest by research groups to the final journal publication. They also include the generation of the appropriate author list and processing of proofs. In addition, metadata are continuously maintained, and public web pages and all the related information regarding the scientific results of the experiment are obtained automatically.

The tools are used by the whole collaboration and minimize the amount of manual work required for repetitive procedures, easing the workload of editors, editorial boards, management, and the Physics Office. At the same time, all documents connected to an analysis can now be accessed from a central tool where the experiment's rules and knowledge are codified and made available in an intuitive way.

The FENCE framework and its integration with the most up to date and efficient tools has consequently provided a more professional and efficient automatized work environment to the entire collaboration.

### Acknowledgements

The authors are indebted to the ATLAS Collaboration for the support provided to achieve the results described in this paper. We are grateful to ATLAS collaborators who provided invaluable comments and input to the paper and the framework it presents. Special acknowledgements go to Marzio Nessi for helping initiate the Glance project in ATLAS and for supporting its development, and to Kathy

Pommes for supervising the Glance team at CERN. Special thanks to Giordon Stark and Zachary Marshall for thoroughly reviewing this paper.